\documentclass[two column]{revtex4}

 \newcommand{\ed}{\end{document}}
 \newcommand{\be}{\begin{equation}}
 \newcommand{\ee}{\end{equation}}
 \newcommand{\er}{\end{eqnarray}}
 \newcommand{\ea}{\end{eqnarray}}
 \newcommand{\br}{\begin{eqnarray}}
 \newcommand{\ba}{\begin{eqnarray}}
 \newcommand{\dslash}{\partial\!\!\!/}
 \newcommand{\aslash}{A\!\!\!/}

\begin{document}

 \title{Conflict between symmetries: a classical example}

 \author{Davi Cabral Rodrigues and Clovis Wotzasek}
 \affiliation{Instituto de F\'\i sica, Universidade
 Federal do Rio de Janeiro\\21945, Rio de Janeiro, Brazil\\}

 \begin{abstract}
 \noindent We have found a (classical) competition between duality and gauge symmetries when trying to obtain an explicit dual to the non-nonabelian version of the self-dual model proposed by Townsend, Pilch and van Nieuwenhuizen\cite{TPvN} (NASD) either through gauge embedding procedures or gauge invariant master approach.  We found that the theory dual to NASD {\it is not} the Yang-Mills-Chern-Simons model. We then proved that a model gauge invariant and dual equivalent to NASD cannot be achieved because of this conflict. Other nonabelian self-dual formulation proposed in \cite{KLRvN} was studied and its dual obtained. We discuss the consequences of the dual equivalence found here in the context of 3D non-abelian bosonization.
\end{abstract}

 \maketitle

 This paper is devoted to study duality mapping and model equivalence in the context of three dimensional nonabelian self-dual models \cite{TPvN}.  There is a widespread belief that the nonabelian version of the model proposed in \cite{TPvN} is equivalent to the topologically massive Yang-Mills-Chern-Simons model (YMCS) proposed in \cite{annals}, the reasons being that (i) their abelian counter-parts are equivalent \cite{DJ} and (ii) the indications of perturbative equivalence when the coupling constant is very small \cite{BFMS,BBG}. This is important for the bosonization program in higher dimensions that aims to establish a connection between the low energy sector of a theory of massive self-interacting, G-charged fermions -- the massive Thirring Model in 2 + 1 dimensions -- into a gauge theory, the YMCS model, at least in that long wavelength limit. In fact some results in this direction have been reported already in the small coupling constant limit \cite{BFMS,BBG}.  We shall establish an exact result showing that these two models {\it are not} equivalent.  We will then find explicitly the dual equivalent for the nonabelian self-dual model and discuss its differences and similarities with the YMCS model.  As a bonus we also discuss another massive self-dual model proposed in \cite{KLRvN} and find the dual equivalent.

 Bosonization was developed in the context of the two-dimensional field theory and has been one of the main tools available to investigate the non-perturbative behavior of some interactive field theories \cite{2dbos}.
 For some time this concept was thought to be exclusive of two-dimensional space-times where spin is absent and one cannot distinguish between bosons and fermions.
 For higher dimensions, due to the absence of an operator mapping a la Mandelstan, the situation is more complicate and even the bosonization identities extracted from these procedures relating the fermionic current with the bosonic topological current is a consequence of a non-trivial current algebra.  Moreover, contrary to the two dimensional case, in dimensions higher than two there are no exact results with the exception of the current mapping \cite{LMNS,LG}.
 In spite of some difficulties, the bosonization program has advanced to higher dimensions \cite{luscher,marino}.
 It was only recently that this powerful technique were extended to higher dimensional space-times following diverse orientations \cite{BLQ,FGM} and \cite{FS,B}.
 The bosonization mapping in D=3, shows that this is a relevant issue in the context of transmutation of spin and statistics in three dimensions \cite{P}.
 The equivalence of the three dimensional effective electromagnetic action of the $CP^1$ model with a charged massive fermion to lowest order in inverse (fermion) mass has been proposed by
 Deser and Redlich \cite{DR}.  Using their results bosonization was extended to three dimensions in the 1/m expansion \cite{FS}. These endeavor has led to promising results in diverse areas such as, for instance, the understanding of the universal behavior of the Hall conductance in interactive fermion systems \cite{BOS} very important to understand transport properties \cite{RSS} and to compute the quark-antiquark potential in three dimensional massive quantum electrodynamics \cite{AB}. In fact the number of applications is quite extensive.

 In particular, the $2+1$ dimensional abelian massive Thirring model (MTM) has been bosonized to a free vectorial theory in the leading order of the inverse mass expansion.
 Using the well known equivalence between the self-dual \cite{TPvN} and the topologically massive models\cite{annals} proved in a seminal paper by Deser and Jackiw \cite{DJ} through the master action approach, a correspondence has been established between the partition functions for the MTM and the Maxwell-Chern-Simons (MCS) theories \cite{FS}.  The situation for the case of fermions carrying non-nonabelian charges, however, is less understood due to a lack of equivalence between these vectorial models, which has only been established for the weak coupling regime \cite{BFMS,BBG}.  The use of master actions in this situation was critically observed in \cite{KLRvN} and \cite{BBG} as being ineffective for establishing dual equivalences.

 In this paper we intend to fill up this gap.  We use a new technique \cite{AINRW,IW} proposed to perform duality mappings for vectorial models in any dimensions that is alternative to the master action approach \cite{GMdS}. Once the two equivalent dual models are found, a master action fitting both ends of the duality is obtained by field integration. This technique \cite{AINRW,IW} is based on a two-fold approach that simultaneously lift a global symmetry in its local form and may be realized by an iterative embedding of Euler vectors counter-terms.  The use of Euler vectors is done to guarantee the dynamical equivalence between the models, while the embedding algorithm progressively subtracts the gauge offending terms from the theory \cite{CW}.  It seems however that for the nonabelian case the lifting to a local symmetry conflicts with dual equivalence and either one must be sacrificed. In this work we will look for duality equivalence in detriment of gauge symmetry.  In another contribution we followed the opposite route and construct a gauge invariant theory out of the gauge variant model \cite{IW2}.  This situation is well known in quantization processes but, as far as we know, is new in classical domains.  The bosonization of the chiral electrodynamics in 2D illustrate the quantum feature -- while in the bosonized model vectorial model an adjustment of the regularization parameter is enough to establish gauge symmetry, such a simple task is not possible in the chiral version \cite{JR}.  We will show below that for NASD even though we subtract the gauge offending term from the original theory, the use of Euler kernels to guarantee duality equivalence will bring another gauge variance into the game.

 First, let us discuss the general idea in \cite{AINRW,IW} and consider the duality mapping from a general gauge variant model described by a Lagrangian density ${\cal L}_0$ and call its dual equivalent as $\mbox{}^*{\cal L}$.
 As shown in \cite{AINRW}, the final effect of the gauge embedding algorithm is materialized in the following form
 \be
 \label{p10}
 {\cal L}_0[A] \to \mbox{}^*{\cal L}[A] = {\cal L}_0[A] + f(K^2)
 \ee
 where the Euler kernel 
 \be
 \label{p20}
 \delta{\cal L}_0[A] = K_\mu[A]\, \delta A^\mu
 \ee
 defines the classical dynamics of the original theory.  Therefore equivalence will be impacted if the final function is demanded to satisfy the condition
 \be
 \label{p30}
 f(K^2)\mid_{K=0} = 0
 \ee
 This technique was originally explored in the context of the
 soldering formalism \cite{ABW,BW} and explored to study equivalent dynamics of vector models in diverse regimes that includes non-linearities and couplings to dynamical matter. It has also been used to investigate the Lorentz-violating
electrodinamics \cite{jose}. It is exploited here since it seems to be the most appropriate technique for non-nonabelian generalization of the dual mapping concept. In general the algorithm works by demanding gauge invariance but we will relax this condition here and choose a simple function in (\ref{p10}) just by inspection.

 To show the full power of the methodology proposed here we illustrate the technique by computing the dual equivalent to a model proposed in \cite{KLRvN} that could not be found by the traditional means,
 \be
 \label{NBB10}
 S_{SD}[A] = \int\!\! d^3 x \left[\frac m2 B_\mu^a B^{\mu a} - \frac 12 \epsilon^{\mu\nu\lambda} B_\mu^a D_{\mu\nu}^a[B]\right]
 \ee
 where
 \be
 D_{\mu\nu}^a[B] = \partial_\mu B_\nu^a +V^{ab}_\mu B_\nu^b
 \ee
 Following the steps described above we compute the (square of) Euler kernel for this model and subtract from the gauge non-invariant piece in order to eliminate the mass term that is the one offending gauge symmetry in this case.  We obtain,
 \be
 \mbox{}^* S[B] = \int\!\! d^3 x \left[- \frac 14 {\cal F}_{\mu\nu}^a{\cal F}^{\mu\nu a} + \frac m2 \epsilon^{\mu\nu\lambda} B_\mu^a D_{\mu\nu}^a[B]\right]
 \ee
 where
 \be
 {\cal F}_{\mu\nu}^a = \partial_{[\mu}B_{\nu]}^a + V_\mu^{ab} B_\nu^b - V_\nu^{ab} B_\mu^b
 \ee
 This action is dual to (\ref{NBB10}) as a direct computation shows but is not gauge invariant which displays the conflict between the symmetries we are discussing -- we either have duality correspondence, as above, or we have gauge symmetry. For the latter we need to add extra terms that will destroy the duality.

 To set the problem of NASD duality in its proper context let us, in the sequence, review the problem of  identifying a bosonic
 equivalent of a three dimensional theory of self-interacting fermions with symmetry group G and show how it is possible to bosonize the low-energy regime of the theory.  To begin with we define the $G$-current,
 \be
 \label{NB10}
 j^{a\mu} = \bar\psi^i t^a_{ij} \gamma^{\mu} \psi^j\, ,
 \ee
 where $\psi^i$ are N two-component Dirac spinors in the fundamental
 representation of $G$, $i,j = 1,\ldots , N$ and $a= 1,\ldots ,\mbox{dim}\: G$.
 Here $t^a$ and $f^{abc}$ are the generators  and the structure constants of the symmetry group $G$, respectively and $j^{a\mu}$ is a $G$-current.

 The fermionic partition function for the three-dimensional massive
 Thirring model is,
 \be
 \label{BB10}
 {\cal Z}_{Th} = \int \,{\cal{D}}\bar\psi {\cal{D}}\psi\; e^{i\int \left ( \bar\psi^i (i\dslash + m) \psi^i -\frac{g^2}{2N} j^{a\mu}j_{\mu}^a
 \right ) d^3x}
 \ee
 with the coupling constant $g^2$ having dimensions of inverse mass.
 Next we eliminate the quartic interaction through a Legendre transformation
 \be
 e^{-\frac{ig^2}{2N} \int\!d^3\!x\, j^{a\mu}j_{\mu}^a } = \int{\cal D} A_{\mu}
 e^{i\int\!d^3\!x\, (\frac{N}{2g^2}A^{a\mu}A_{\mu}^a + j^{a\mu}A_{\mu}^a)}
 \label{NB20}
 \ee
 up to a multiplicative normalization constant, where we have introduced a vector field
 $A_{\mu}$ taking values in the Lie algebra of G.  After integration of the fermionic degrees of freedom, the partition function reduces to,
 \be
 {\cal Z}_{Th} =\int {\cal{D}} A_{\mu}
  \det (i\dslash + m + \aslash) \;e^{i\frac{N}{2g^2}\!\int d^3 x\:   \left(A^{a\mu}A^a_{\mu}\right)}  .
 \label{NB30}
 \ee

 The determinant of the Dirac operator is an unbounded operator and requires regularization.
 For D=2 this determinant can be computed exactly, both for abelian and non-abelian symmetries.
 Based on general grounds only, one may say that this determinant consists of a Chern-Simons action standing as the leading term plus an infinite series of terms depending on the dual of the vector field, $\tilde F_\mu \sim \epsilon_{\mu\nu\lambda}\partial^\nu A^\lambda$, including those terms that are non-local and non-quadratic in $\tilde F_\mu$.
 For the D=3 the actual computation of this determinant will give parity breaking and parity conserving terms that are computed in powers of the inverse mass,
 \be
 \ln \det (i\dslash + m + \aslash) =   \frac{i\chi}{16\pi} {\cal S}_{CS}[A]
 + i I_{PC}[A] + O(\frac{\partial^2}{m^2}) .
 \label{NB40}
 \ee
 Here ${\cal S}_{CS}$ given by
 \be
 \label{NB50}
 {\cal S}_{CS}[A] = - \int d^3 x\:  \epsilon^{\mu\nu\lambda}\, tr 
 (F_{\mu \nu} A_{\lambda} - \frac{2}{3} A_{\mu}A_{\nu}A_{\lambda})  ,
 \ee
 is the non-abelian Chern-Simons action and the parity conserving contributions, in first-order, is the Yang-Mills action
 \be
 I_{PC}[A] =  \frac{1}{24\pi  m}\, tr\int d^3 x\: F^{\mu\nu} F_{\mu\nu}  ,
 \label{NB60}
 \ee
 where
 \be
 F_{\mu\nu} = \partial_{\mu}A_{\nu} - \partial_{\nu}A_{\mu} + [A_{\mu},A_{\nu}]  .
 \label{NB70}
 \ee

 In the low energy regime, only the Chern-Simons action survives yielding a closed expression for the determinant as,
 \be
 \lim_{m\to\infty} \det (i\dslash + m + \aslash) =   \frac{i\chi}{16\pi} {\cal S}_{CS}[A]
 \label{NB75}
 \ee
 Using this result we can write ${\cal Z}_{Th}$ in the form
 \be
 {\cal Z}_{Th} = \int {\cal{D}} A_{\mu}  \exp(i{\cal S}_{SD}[A])  ,
 \label{NB80}
 \ee
 where ${\cal S}_{SD}$ is the non-abelian version of the self-dual action  introduced in \cite{TPvN},
 \ba
 \label{pp220}
 \!\!\!\!\!{\cal S}_{SD}[A] = -\frac N{g^2}\int d^3 x\: tr \left(A_\mu A^\mu \right) - \frac \chi{8\pi} {\cal S}_{CS}[A]\nonumber\\
 \!\!\!\!= \!\!\int\!\! d^3 x\!\!\left[\! \frac N{2g^2}  A^a_\mu A^{a\mu}\!\! +\!\! \frac \chi{8\pi} \epsilon^{\mu\nu\lambda}\!\!\!
 \left(\!\!A_\mu^a \partial_\nu A_\lambda^a\!\!+\!\!\frac {f^{abc}}{3}A_\mu^a A_\nu^b A_\lambda^c \!\!\right)\!\!\right]
 \ea
 Therefore, to leading order in $1/m$ we have established the identification ${\cal Z}_{Th} \approx  {\cal Z}_{SD} $.
 It is interesting to observe that the Thirring coupling constant
 $g^2/N$ in the fermionic model is mapped into the inverse mass spin $1$ massive excitation,
 $\mu = 4\pi N/g^2$.

 We are now in position to prove our main result -- the actual computation of the dual equivalent to the self-dual model (\ref{pp220}).  To do this we need to compute the Euler vector to the model
 \be\label{PP230}
 K_\mu^a = \frac N{g^2}  A^a_\mu  + \frac \chi{8\pi} \epsilon^{\mu\nu\lambda}
 F_{\nu\lambda}^a
 \ee
 and choose a function $f(K^2)$ so that the gauge offending term in (\ref{pp220}) gets subtracted out.  It is simple to see that the following choice will do,
 \ba
 \label{pp240}
 &&\mbox{}^*{\cal S}[A] \!=\! {\cal S}_{SD}[A] - \frac {g^2}{2 N} \int d^3 x\: K_\mu^a K^{\mu a}\nonumber\\
 &=& \!\!\!\frac 1{8\pi}\!\! \int\!\! d^3 x\! \left\{\! -\frac {g^2}{8\pi N} F_{\mu\nu}^a F^{\mu\nu a}\right.\nonumber\\
 &-& \left.\chi \epsilon^{\mu\nu\lambda}\!\!\left(\!A_\mu^a \partial_\nu A_\lambda^a\!\!+\!\!\frac 23{f^{abc}}A_\mu^a A_\nu^b A_\lambda^c\!\right) \! \right\}
 \ea
 leading to a Yang-Mills type theory.  Notice that the second term is Chern-Simons like term but not quite the CS itself.  By construction both theories have the same dynamics but the dual model is still not gauge invariant. Although we start by subtracting the gauge variant piece, the counter-term added has broken the CS term spoiling its gauge invariance for small gauge transformations. To produce a gauge invariant model, at least for those transformations connected to the identity, one has to add a counter-term not proportional to the Euler kernel which will eventually spoil dynamical equivalence.  It becomes then quite clear that the YMCS model is not dual to the nonabelian self-dual model.  The model in (\ref{pp240}) is and this is an exact result.

 Once the duality is established between actions (\ref{pp220}) and (\ref{pp240}) we may try to find a master action ${\cal S}_{M}[A,B]$ (certainly not a gauge invariant one) that will fit both ends when variations are taken with respect to $B$ and $A$ fields, respectively. After some experimentation we have found that the following acion will produce the desired results
\ba
 \label{pp260}
 \!\!\!\!\! {\cal S}_{M}[A,B] && = \int d^3x \left [ \frac N {2g^2} B^{a\mu} B^a_\mu + \frac \chi {8\pi} \epsilon^{\mu \nu \lambda} B_\mu^a F^a_{\nu \lambda} (A) - \right. \nonumber \\
	&& \left. - \frac \chi {8\pi} \epsilon^{\mu \nu \lambda} \left (A^a_\mu \partial_\nu A^a_\lambda + \frac 23 f^{abc} A_\mu^a A^b_\nu A^c_\lambda \right ) \right ]
\ea


 Now comes an important observation.  Unlike the master approach, our result, the action $\mbox{}^*{\cal S}[A]$ in (\ref{pp240}), is dual to NASD for all values of the coupling constant.  This is not the YMCS model.  This is a new and very important result since all previous investigations, either perturbative \cite{BFMS} or Stuckelberg embedding \cite{BBG}, could not point to a clear direction. The proof, based on the use of a gauge invariant ``interpolating Action''  is seen to run into trouble in the non-abelian case. That the non-abelian extension of this kind of equivalences is more
 involved was already recognized in \cite{DJ} and \cite{BFMS,BBG}, and shown that the non-abelian self-dual
 action {\it is not}
 equivalent to a Yang-Mills-Chern-Simons theory (the natural extension
 of the abelian MCS theory) but to a model where the
 Yang-Mills term vanishes in the limit $g^2 \to 0$ \cite{BFMS}. In this paper we propose an explicit candidate for this vacuous, lasting for over a 20 years period.
Finally, let us comment on a previous contribution in this direction \cite{IW2}.  In that paper we failed to notice the incompatibility reported here.  As a consequence an extra term was added to the answer to establish gauge invariance.  As shown above this step destroyed the duality in favor of gauge symmetry.  We take this opportunity to correct that result.

 \end{document}